\documentclass[twocolumn]{aastex701}

\received{XXX}
\revised{YYY}
\accepted{ZZZ}

\submitjournal{ApJ}

\shorttitle{Radial velocity evidence for NaSt1}
\shortauthors{Wangnok et al.}

\begin{document}

\title{Radial Velocity Evidence for a Post-Mass-Transfer Massive Binary System NaSt1}

\correspondingauthor{Samaporn~Tinyanont}
\email{samaporn@narit.or.th}

\author[0000-0002-7593-2748]{Kittipong~Wangnok}
\affiliation{School of Physics, Institute of Science, Suranaree University of Technology, Nakhon Ratchasima 30000, Thailand}
\affiliation{National Astronomical Research Institute of Thailand, 260 Moo 4, Donkaew, Maerim, Chiang Mai, 50180, Thailand}
\email{kittipong@narit.or.th} 

\author[0000-0002-1481-4676]{Samaporn~Tinyanont}
\affiliation{National Astronomical Research Institute of Thailand, 260 Moo 4, Donkaew, Maerim, Chiang Mai, 50180, Thailand}
\email{samaporn@narit.or.th} 

\author[0000-0003-0778-0321]{Ryan~M.~Lau}
\affiliation{IPAC, Mail Code 100-22, Caltech, 1200 E. California Blvd. Pasadena, CA 91125, USA}
\email{ryanlau@ipac.caltech.edu}

\author[0000-0002-2445-5275]{Ryan~J.~Foley}
\affiliation{Department of Astronomy and Astrophysics, University of California, Santa Cruz, CA 95064, USA}
\email{foley@ucsc.edu}


\author[0000-0003-1305-3761]{R.~Paul~Butler}
\affiliation{Earth \& Planets Laboratory, Carnegie Institution for Science, 5241 Broad Branch Road NW, Washington, DC 20015, USA}
\email{paul@dtm.ciw.edu}

\author[0000-0002-9099-4613]{Poemwai~Chainakun}
\affiliation{School of Physics, Institute of Science, Suranaree University of Technology, Nakhon Ratchasima 30000, Thailand}
\affiliation{Center of Excellence in High Energy Physics and Astrophysics, Institute of Science, Suranaree University of Technology, Nakhon Ratchasima, 30000, Thailand}
\email{pchainakun@sut.ac.th}

\author[0000-0002-5680-4660]{Kyle~W.~Davis}
\affiliation{Department of Astronomy and Astrophysics, University of California, Santa Cruz, CA 95064, USA}
\email{kywdavis@ucsc.edu}

\author[0000-0003-1481-8076]{Kishalay~De}
\affiliation{Department of Astronomy, Columbia University, 550 W 120th Street, New York, NY 10027, USA}
\affiliation{Center for Computational Astrophysics, Flatiron Institute, 162 Fifth Avenue, New York, NY 10010, USA}
\email{kde3038@columbia.edu}

\author[0000-0003-1012-3031]{Jared~A.~Goldberg}
\affil{Department of Physics and Astronomy, Michigan State University, East Lansing, MI 48824, USA}
\affiliation{Center for Computational Astrophysics, Flatiron Institute, 162 Fifth Avenue, New York, NY 10010, USA}
\email{goldstar@msu.edu}

\author[0000-0002-6153-3076]{Bradford~Holden}
\affiliation{UCO/Lick Observatory, Department of Astronomy and Astrophysics, University of California at Santa Cruz, Santa Cruz, CA 95064, USA}
\email{holden@ucolick.org}

\author[]{Adriana~Kuehnel}
\affiliation{Earth \& Planets Laboratory, Carnegie Institution for Science, 5241 Broad Branch Road NW, Washington, DC 20015, USA}
\email{akuehnel@carnegiescience.edu}

\author[0000-0001-5510-2424]{Nathan~Smith}
\affiliation{Steward Observatory, University of Arizona, Tucson, AZ 85721, USA}
\email{nathans@as.arizona.edu}

\author[0000-0002-9486-818X]{Jonathan~J.~Swift}
\affil{The Thacher School, 5025 Thacher Rd., Ojai, CA 93023, USA}
\email{jswift@thacher.org}

\author[0000-0001-7177-7456]{Steven~S.~Vogt}
\affiliation{UCO/Lick Observatory, Department of Astronomy and Astrophysics, University of California at Santa Cruz, Santa Cruz, CA 95064, USA}
\email{vogt@ucolick.org}









\begin{abstract}

We present multi-epoch high-resolution optical spectroscopy ($R \simeq 80{,}000$) of the emission-line object NaSt1 to test its proposed binary nature, along with long-term multiband photometry, mid-infrared spectroscopy, and spatially resolved integral field unit (IFU) spectroscopy to probe the circumstellar kinematics of the system. 
We detect two groups of 36 emission lines showing radial velocity (RV) variation with a mean period of 311 $\pm$ 5 d, but varying in opposite phase. We associate these two groups with the optically thick wind of the stripped primary star and the wind-wind collision region with the companion star, providing strong evidence for binarity.
The RV and light curve (LC) periods are consistent within the uncertainties, ruling out ellipsoidal modulation, which would require an orbital period of about 620 d. 
The RV–LC phase relationship and high-ionization lines favor binary interaction over pulsations.
We model the 1--5~$\mu$m spectrum of NaSt1 and find two optically thin dust components: hot $T_{\rm h} \simeq 1230$ K, $M_{\rm h} \simeq 2 \times 10^{-10} M_{\odot}$ and warm $T_{\rm c} \simeq 660$ K, $M_{\rm c} \simeq 3 \times 10^{-8} M_{\odot}$. IFU spectroscopy spatially resolves the circumstellar medium in the [\ion{N}{2}] $\lambda6548$ and $\lambda6584$ emission lines, showing a deprojected expansion velocity of $\sim31$ km~s$^{-1}$, implying a dynamical age of $\sim40$ yr. 
This short timescale suggests that the nebula was produced by recent mass loss. The system may represent a Galactic analog of a massive binary undergoing a mass-loss process to become a stripped-envelope supernova progenitor.

\end{abstract}


\section{Introduction} \label{sec:intro}

NaSt1 was first cataloged by \cite{Nassau_1965} as an emission-line source (IRAS 18497+0056; WR 122). NaSt1 has been suggested to be a massive binary system that has recently undergone mass transfer, based on its nitrogen-enriched circumstellar nebula and equatorial morphology \citep[e.g.,][]{Smith_2011RY,Mauerhan_2015}. Spectroscopic studies by \cite{Crowther_1999} revealed a highly ionized nebula dominated by [\ion{N}{2}] and [\ion{N}{3}] emission lines, with depleted carbon indicating CNO-processed material. They also noted that the emission lines are significantly narrower than those observed in typical Wolf-Rayet (WR) stars, whose stellar winds reach velocities of $\sim10^3$~km~s$^{-1}$ \citep{Crowther_2007}. This difference suggests that the dominant line-forming region in NaSt1 differs from that of a classical WR stellar wind, in which emission lines emerge from a fast, optically thick outflow. NaSt1, along with RY Scuti \citep{Smith_2011RY,Smith_2002}, may represent rare Galactic systems currently in, or just past, a brief mass-transfer phase. In both systems, the primary star appears to be massive and to have lost most or all of its hydrogen envelope. If such a star were to explode, it would likely produce a stripped-envelope supernova (SESN). Systems such as NaSt1 offer a rare opportunity to directly probe the mass-loss processes that govern massive-star evolution. 

Mass loss governs the evolution of massive stars, yet the dominant envelope-removal mechanisms remain uncertain. Early models relied on steady line-driven winds to form WR stars, but clumping-corrected mass-loss rates are lower and reduce the efficiency of this channel. Observations instead show high binary fractions \citep{Sana_2012} and widespread eruptive or interaction-driven mass loss, indicating a major role for binary evolution in producing stripped stars and their supernova progenitors \citep[see][for a review]{Smith_2014}. These uncertainties affect predictions of stellar feedback, chemical enrichment, and compact-remnant formation, challenging traditional single-star evolutionary pathways.

X-ray observations provide an additional probe of the physical conditions in NaSt1. NaSt1 was observed using Chandra and the Advanced CCD Imaging Spectrometer (ACIS), revealing a faint X-ray point source coincident with the optical and infrared (IR) position. The emission is dominated by 1--2 keV photons and is strongly absorbed below 1 keV, resembling the WN star WR 18 and consistent with an optically thin thermal plasma typical of WR stars \citep{Mauerhan_2015,Skinner_2010}. Using an absorption column inferred from the optical extinction gives an unabsorbed luminosity of $L_{\rm X} \sim 10^{32}$--$10^{33} \mathrm{erg~s^{-1}}$ and $L_{\rm X}/L_{\mathrm{bol}} \sim 10^{-7}$--$10^{-5}$, within the range of single WR stars and colliding-wind binaries \citep{Skinner_2010,Gagne_2010}. No prominent hard X-ray component is detected, possibly due to limited sensitivity, slow winds, or orbital variability \citep{Luo_1990,Usov_1992}. These properties are consistent with shock-heated plasma from massive-star winds and support a WR system that may involve binary interaction \citep{Mauerhan_2015}. 

The circumstellar medium (CSM) around NaSt1 is spatially resolved in high-resolution imaging with the Hubble Space Telescope (HST) and ground-based infrared adaptive optics (AO) observations \citep{Mauerhan_2015}. 
HST narrow-band imaging revealed a flattened, disk-like nebula in the [\ion{N}{2}] emission line, while the continuum source remains unresolved \citep{Mauerhan_2015}. This morphology is consistent with materials recently ejected during binary interaction, although the system’s binarity has not yet been confirmed. The optical and IR emission likely arises from the scattered and reprocessed light from the embedded central source. Radio observations from the Galactic Plane Survey with the Karl G. Jansky Very Large Array (GLOSTAR; \citealp{Medina_2019}) detected thermal free-free emission consistent with a dense, ionized stellar wind. Using the radio flux-mass-loss relation of \cite{Wright_1975}, the inferred mass-loss rate is about $10^{-4} M_{\odot}\,\mathrm{yr}^{-1}$. This value lies at the extreme upper end of rates for steady winds observed in evolved massive stars, such as WR stars \citep{Crowther_2007} and extreme red supergiants \citep{Smith_2014}. 


NaSt1 represents a rare opportunity to directly observe a massive binary system during or shortly after Roche-lobe overflow (RLOF) and non-conservative mass transfer. Despite the importance of binary interaction in the evolution of massive stars, systems caught in these brief evolutionary stages remain extremely rare. Massive stars themselves constitute only a small fraction of the Galactic stellar population owing to the initial mass function and their short lifespan. Combined with the short duration of the RLOF phase ($\sim10^4$ yr; \citealp{Podsiadlowski_1992}), which corresponds to only $\sim10^{-3}$ of a typical massive star lifetime ($\sim10^7$ yr; \citealp{Ekstrom_2012}), the probability of observing systems in this evolutionary stage is low. NaSt1 and RY Scuti are the only two known systems with resolved CSM. For RY Scuti, the CSM tori have dynamical ages of only $\sim120$ and ~250 yr \citep{Smith_2011RY}, indicating that the system is being observed shortly after major mass-ejection events. Similarly, the compact nebula and rich emission-line spectrum of NaSt1 suggest ongoing or very recent interaction between the binary and its circumstellar environment.


Photometric light curve (LC) monitoring from the Palomar Gattini-InfraRed (PGIR; \citealp{De_2020a,De_2020b}) survey, the Zwicky Transient Facility (ZTF; \citealp{Bellm_2019,Masci_2019}), the All-Sky Automated Survey for Supernovae (ASAS-SN; \citealp{Shappee_2014, Kochanek_2017}), the Asteroid Terrestrial-impact Last Alert System (ATLAS; \citealp{Heinze_2018,Tonry_2018,Smith_2020}), and the Katzman Automatic Imaging Telescope (KAIT; \citealp{Filippenko_2001}) revealed periodic brightness variations with a period of $\sim$310 d \citep{Lau_2021}. 




To explain the overall variability, \cite{Lau_2021} suggested a binary origin and discussed several possible mechanisms, including ellipsoidal modulation and circumstellar obscuration (e.g., from colliding winds). Although this behavior is strongly consistent with orbital motion in an enshrouded binary, radial velocity (RV) measurements are required to confirm whether NaSt1 is truly a binary system.


Here we present multi-epoch high-resolution spectroscopy of NaSt1, complemented by photometry, mid-IR spectroscopy, and integral field unit (IFU) observations. The observations are described in Section~\ref{sec:obs}. In Section~\ref{sec:analysis}, we identify 36 emission lines and measure their RVs using the cross-correlation function (CCF). We also derive dust properties from the IR spectrum and present spatially resolved gas kinematics from the IFU data. In Section~\ref{sec:discussions}, we discuss the observed variability by combining the LC and RV measurements. We examine the origin of the $\sim$310 day RV period, the line-forming regions, and the structure of the CSM and dust. We also compare NaSt1 to RY Scuti. Finally, we summarize our results in Section~\ref{sec:conclusions}.

\section{Observations, Data Reduction, and Archival Data} \label{sec:obs}
\subsection{APF Spectroscopy}

We acquired seven epochs of high-resolution spectra of NaSt1 using the Levy spectrometer mounted on the 2.4-m Automated Planet Finder (APF) telescope at Lick Observatory. The Levy spectrometer is a high-resolution slit-fed optical echelle spectrograph designed to measure Doppler velocities with a precision better than a meter per second \citep{Radovan_2010,Vogt_2014}. It provides simultaneous wavelength coverage of 3740--9700~\AA. We used the $2\arcsec \times 8\arcsec$ slit, providing $R \simeq 80{,}000$ at 550~nm. Levy's slit is fixed 19\textdegree\ from parallactic. The spectrograph is equipped with an atmospheric dispersion corrector. The slit was centered on NaSt1 at $\alpha = 18^{\text{h}} 52^{\text{m}} 17.55^{\text{s}}$, $\delta = +00^{\circ} 59' 44.30''$ (J2000). The observation log, including the slit position angle (PA) range for each epoch, is presented in Table~\ref{tab:observations}.

\begin{figure}[!hbt]
\begin{center}
\includegraphics[width=0.45\textwidth]{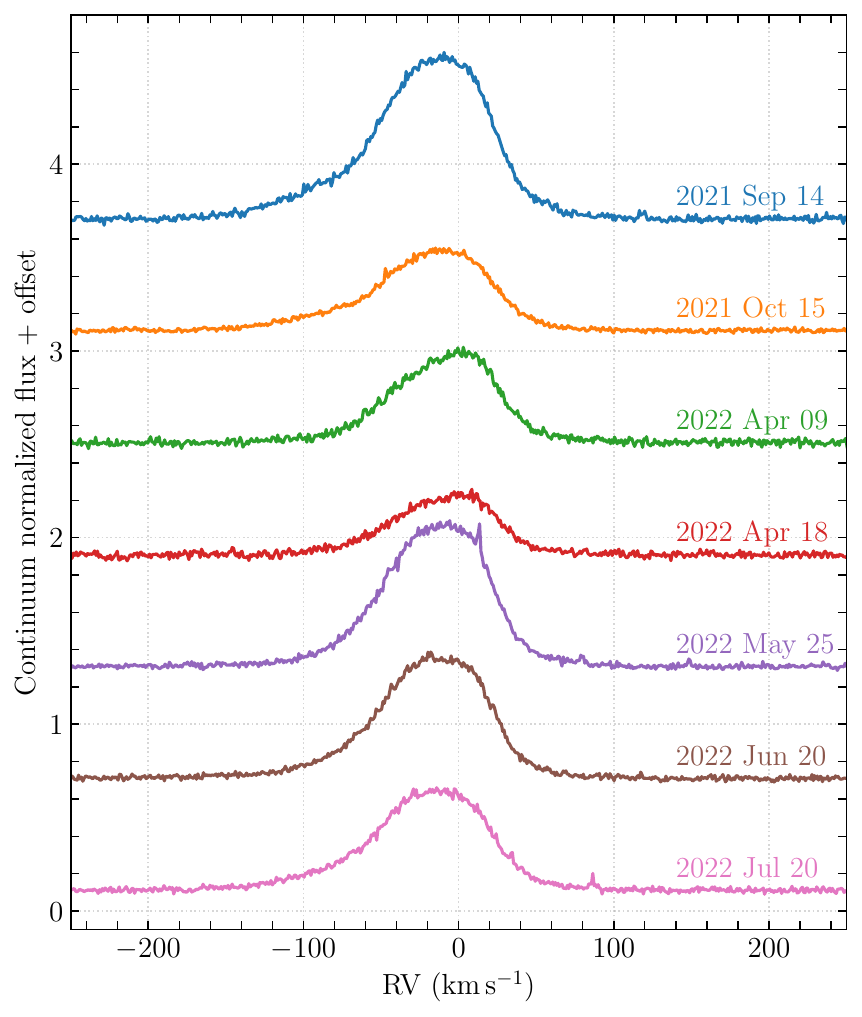}
\caption{High-resolution \ion{He}{2} $\lambda$4686 line profiles of NaSt1 from seven observing epochs between 2021 and 2022. The spectra are normalized to the continuum and corrected for barycentric motion. They are vertically offset for clarity.}
\label{fig:HeII4686_rv}
\end{center}
\end{figure}

The data were reduced using a standard data reduction pipeline, performing echelle order tracing using internal flat observations. 
Wavelength calibration was performed using the iodine cell with nightly observations. ThAr arc lamps were observed to monitor night-to-night variations in the wavelength solution. The instrumental wavelength solution is highly stable, with variations of approximately 0.2 pixels between nights.
An example of \ion{He}{2} $\lambda$4686 obtained from APF observations is presented in Figure \ref{fig:HeII4686_rv}.
Figure~\ref{fig:line_profiles_all} shows 36 continuum-normalized spectra of NaSt1 obtained with the APF over seven observing epochs.

\begin{table*}[!hbt]
\centering
\caption{Log of high-resolution spectroscopic observations of NaSt1.}
\begin{tabular}{cccccc}
\hline
\hline
Epoch Number & UT Date & MJD & Exposure Time (s) & Slit PA Range (deg)\\
\hline
\centering
1 & 2021 Sep 14 & 59471.178 & 8220.6 & $-$2.39 $\rightarrow$ 14.54\\
2 & 2021 Oct 15 & 59502.804 & 4121.5 & $-$2.60 $\rightarrow$ 0.64\\
3 & 2022 Apr 09 & 59678.466 & 4773.0 & 54.68 $\rightarrow$ 73.01\\
4 & 2022 Apr 18 & 59687.407 & 3600.0 & 72.36 $\rightarrow$ 83.97\\
5 & 2022 May 25 & 59724.417 & 7965.2 & 11.61 $\rightarrow$ 48.31\\
6 & 2022 Jun 20 & 59750.386 & 8400.0 & 0.21 $\rightarrow$ 30.55\\
7 & 2022 Jul 20 & 59780.423 & 8400.0 & -2.61 $\rightarrow$ 5.45\\
\hline
\end{tabular}
\label{tab:observations}
\end{table*}

\subsection{Optical to near-infrared photometry}

\begin{figure*}[!hbt]
\begin{center}
\includegraphics[width=1\textwidth]{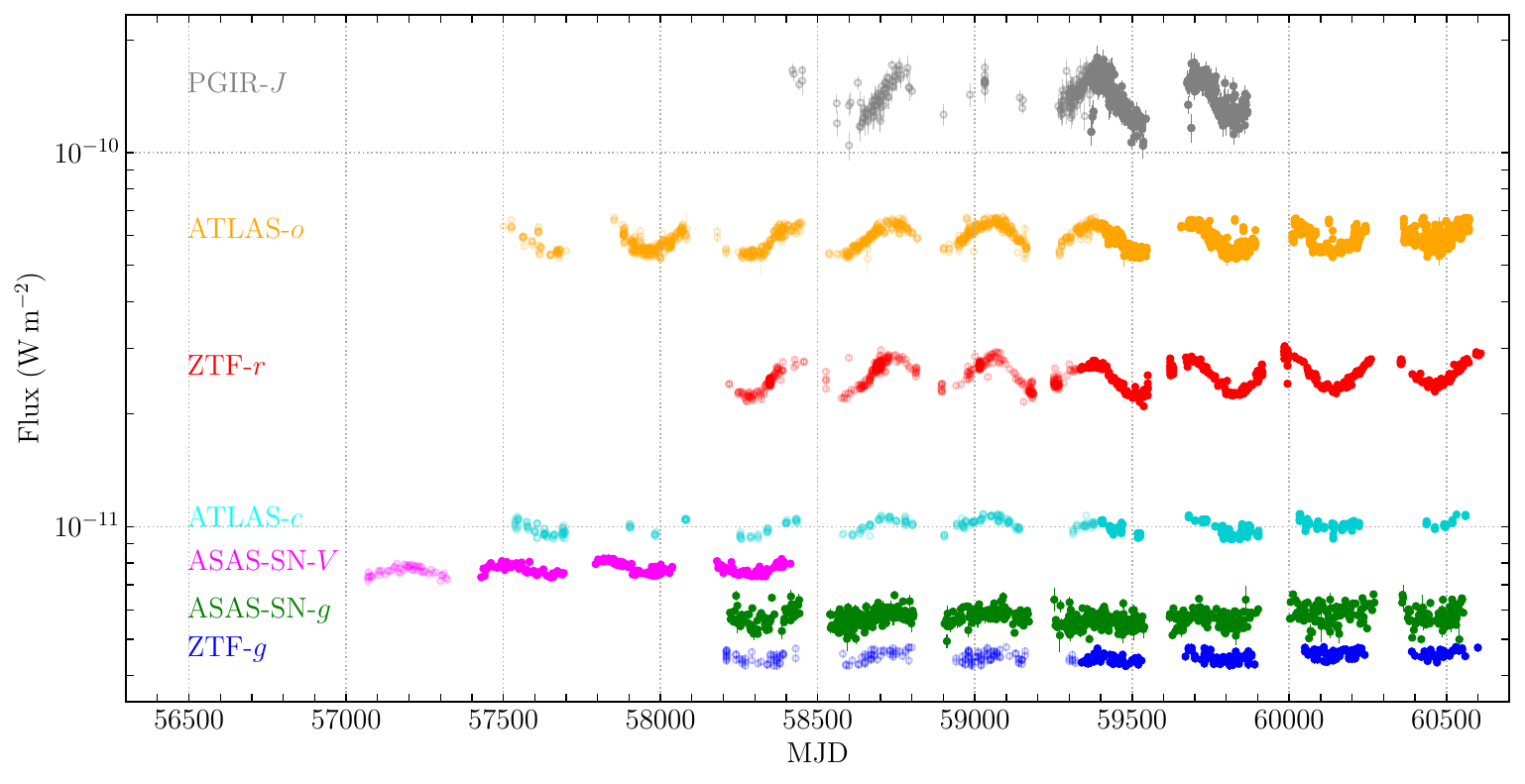}
\caption{Multiband photometric light curves of NaSt1 in ZTF-$g$, ZTF-$r$, ASAS-SN-$g$, ASAS-SN-$V$, ATLAS-$c$, ATLAS-$o$, and PGIR-$J$. Open symbols denote photometry reported in \citet{Lau_2021}, while filled symbols represent archival data analyzed in this work.}
\label{fig:nast1_lc}
\end{center}
\end{figure*}

We obtained public photometry of NaSt1 from the ZTF \citep{Bellm_2019,Masci_2019}, ASAS-SN \citep{Shappee_2014, Kochanek_2017}, ATLAS \citep{Heinze_2018,Tonry_2018,Smith_2020}, and PGIR \citep{De_2020a,De_2020b} in the $J$ band. ZTF data in the $g$ and $r$ bands are available through the NASA/IPAC Infrared Science Archive (IRSA)\footnote{\url{https://irsa.ipac.caltech.edu/Missions/ztf.html}}. ASAS-SN includes  light curves in the $g$ and $V$ bands\footnote{\url{http://asas-sn.ifa.hawaii.edu/skypatrol/}}. NaSt1 was recognized as a variable star with the reference ID ASASSN-V J185217.55+005944.3 in the Variable Star Database \citep{Jayasinghe_2020}. The ATLAS project also provides the NaSt1 light curve in the ATLAS-$c$ ($\lambda_{\mathrm {eff}}$ = 5184 \AA) and ATLAS-$o$ ($\lambda_{\mathrm {eff}}$ = 6632 \AA) bands, which can be accessed through the ATLAS forced-photometry server\footnote{\url{https://fallingstar-data.com/forcedphot/}}. All available LCs of NaSt1 are shown in Figure~\ref{fig:nast1_lc} in ZTF-$g$, ZTF-$r$, ASAS-SN-$g$, ASAS-SN-$V$, ATLAS-$c$, ATLAS-$o$, and PGIR-$J$. The final data points from each survey correspond to the last epochs reported by \citet{Lau_2021}, with MJDs of 59324.4822 (ZTF-$g$), 59329.4404 (ZTF-$r$), 57341.117 (ASAS-SN-$V$), 59402.4035 (ATLAS-$c$), 59404.3875 (ATLAS-$o$), and 59367.2876 (PGIR-$J$). Data obtained after these epochs are newly presented in this work. We also present ASAS-SN-$g$ data, which were not reported previously.

\subsection{SpeX mid-infrared spectroscopy}
We observed NaSt1 with the SpeX spectrograph \citep{rayner2003} on the NASA InfraRed Telescope Facility (IRTF) on 2021 June 21 (UT date used throughout the paper; MJD 59386). 
Observations of an A0V star were obtained close in airmass and time to provide telluric correction. 
We obtained spectra using the $0\farcs3 \times 15\arcsec$ slit in the short cross-dispersed and long cross-dispersed long modes, covering 0.7--2.55 $\mu$m at $R\sim 1200$ and 1.98--5.3 $\mu$m at $R \sim1500$, respectively.
The data were reduced using \texttt{spextool} \citep{cushing2004} and telluric corrected using \texttt{xtellcor} \citep{vacca2003}.

\subsection{Integral field unit spectroscopy}
NaSt1 was observed on 2025 May 26 (MJD 60821), with the Keck Cosmic Web Imager (KCWI; \citealp{Morrissey_2018_KCWI}) on the Keck II telescope a top Mauna Kea in Hawai'i. Observations were taken in both the blue and red channels, providing usable wavelength coverage from $\sim$3500--8000~\AA. The medium slicer was employed in combination with the BL and RL gratings, yielding a spectral resolution of $R \sim 1800$ in the blue and $R \gtrsim 1000$ in the red. We exposed for 120~s in each channel. The field of view was $16\arcsec \times 20\arcsec$ with a spatial sampling of $0.70\arcsec$ per spaxel. The data were reduced and assembled into IFU datacubes using the automated \texttt{KCWI\_DRP} pipeline \citep{Neill_2023_KCWIDRP} and retrieved from the Keck Observatory Archive.

\section{Analysis} \label{sec:analysis}
\subsection{Radial velocity measurements from cross-correlation function}

We measured RV shifts across multiple epochs using a CCF analysis. All spectra were corrected for the Earth's motion to the barycentric reference frame using the Astropy barycentric velocity correction routines \citep{Astropy_2013,Astropy_2018}. The spectra were resampled to a common wavelength grid using linear interpolation (\texttt{numpy.interp}; \citealt{Harris_2020}), normalized to the continuum (fitted using a third order polynomial in each order), and cleaned of outliers using $3\sigma$ clipping (\texttt{FittingWithOutlierRemoval} in \texttt{Astropy}; \citealt{Astropy_2013}). 
We identify 36 emission lines that are clearly detected above the continuum and exhibit well-defined line profiles, as listed in Table~\ref{tab:period_fit}. These lines correspond to those reported by \citet{Crowther_1999}. 
We have identified and analyzed an additional emission line, \ion{He}{1} $\lambda7065$ emission line. This line was not included in the analysis of \citet{Crowther_1999} because it falls within an inter-order gap in their echelle spectrograph.

The profiles of all 36 emission lines are shown in Figure~\ref{fig:line_profiles_all} in the Appendix. The emission-line profiles include both single- and double-peaked structures that arise from different regions of the CSM, as identified by \citet{Crowther_1999}. For each emission line, spectra were converted to velocity space and RV shifts were measured by cross-correlating the profiles in epochs 2--7 with the first epoch. A Gaussian profile was fitted to the peak of each CCF using \texttt{Gaussian1D} in \texttt{Astropy} to determine the velocity shift and its associated uncertainty. This procedure yielded six RV measurements per line. The full orbital solution allowing for eccentricity has six free parameters: the orbital period ($P$), the RV semi-amplitude ($K$), the systemic velocity ($\gamma$), the time of periastron passage ($T_0$), the orbital eccentricity ($e$), and the argument of periastron ($\omega$). However, the current RV dataset consists of only seven measurements per emission line, which is insufficient to constrain an eccentric solution. Therefore, we adopt a circular orbit model (four free parameters). We note that the light curves of NaSt1 are well explained by simple sinusoidal models \citep{Lau_2021}, making a significantly eccentric orbit unlikely.

\begin{figure*}[!hbt]
\begin{center}
\includegraphics[width=0.75\textwidth]{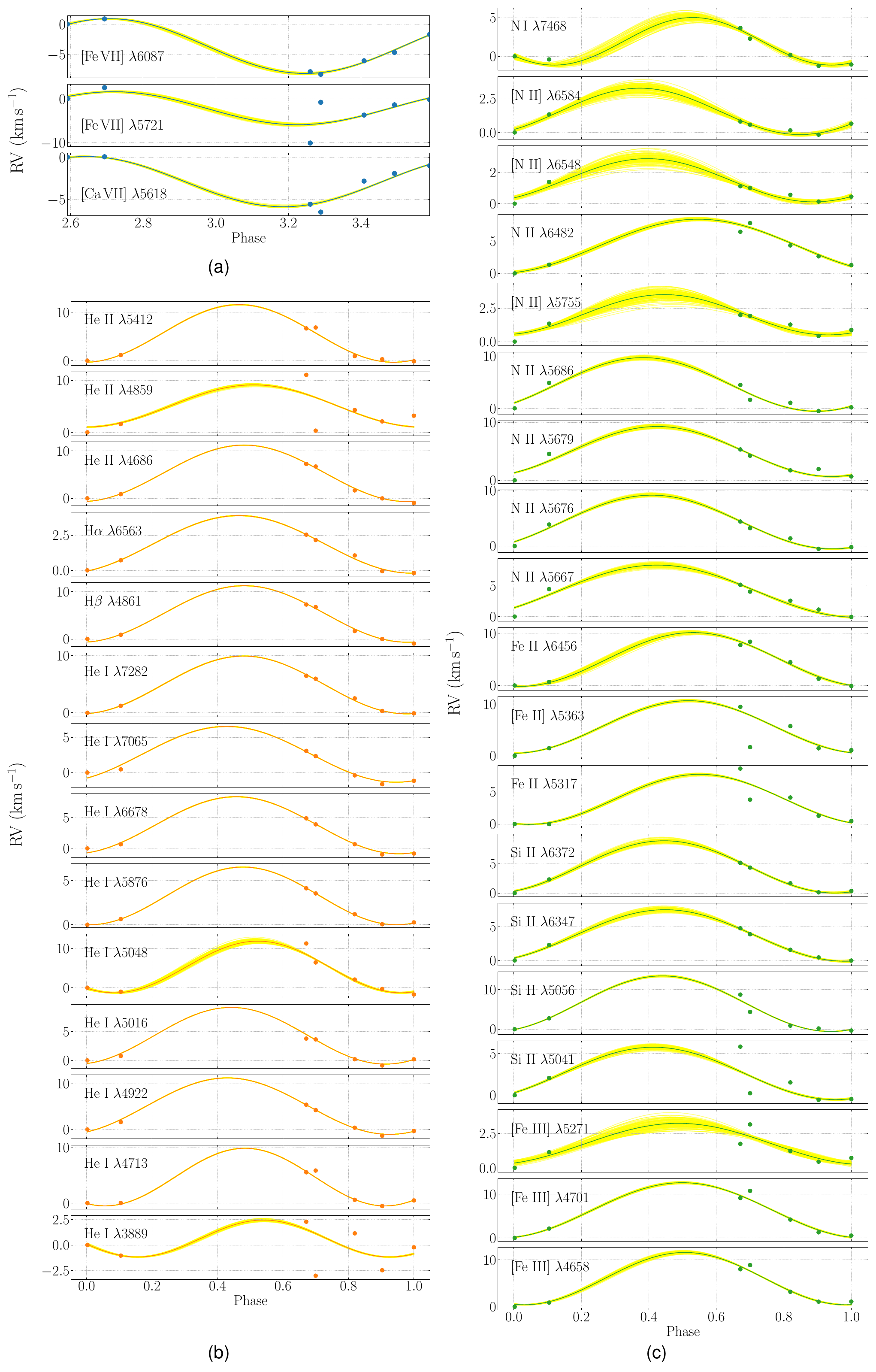}
\caption{RV variations with respect to the first epoch for 36 emission lines in NaSt1 derived from the CCF analysis. Panels (a), (b), and (c) show representative lines associated with the wind-wind collision (WWC) region, the inner circumbinary material (ICM), and the outer circumbinary material (OCM), respectively (see Section~\ref{sec:line_forming}). The highest ionization lines in panel (a) exhibit a phase offset relative to those in panels (b) and (c). The yellow thin curves show 100 randomly selected realizations from the MCMC posterior distributions illustrating the range of radial-velocity variations, while the thick curves show the median posterior model.
}
\label{fig:NaSt1_RV_Variation}
\end{center}
\end{figure*}

\begin{figure*}[!hbt]
\begin{center}
\includegraphics[width=0.925\textwidth]
{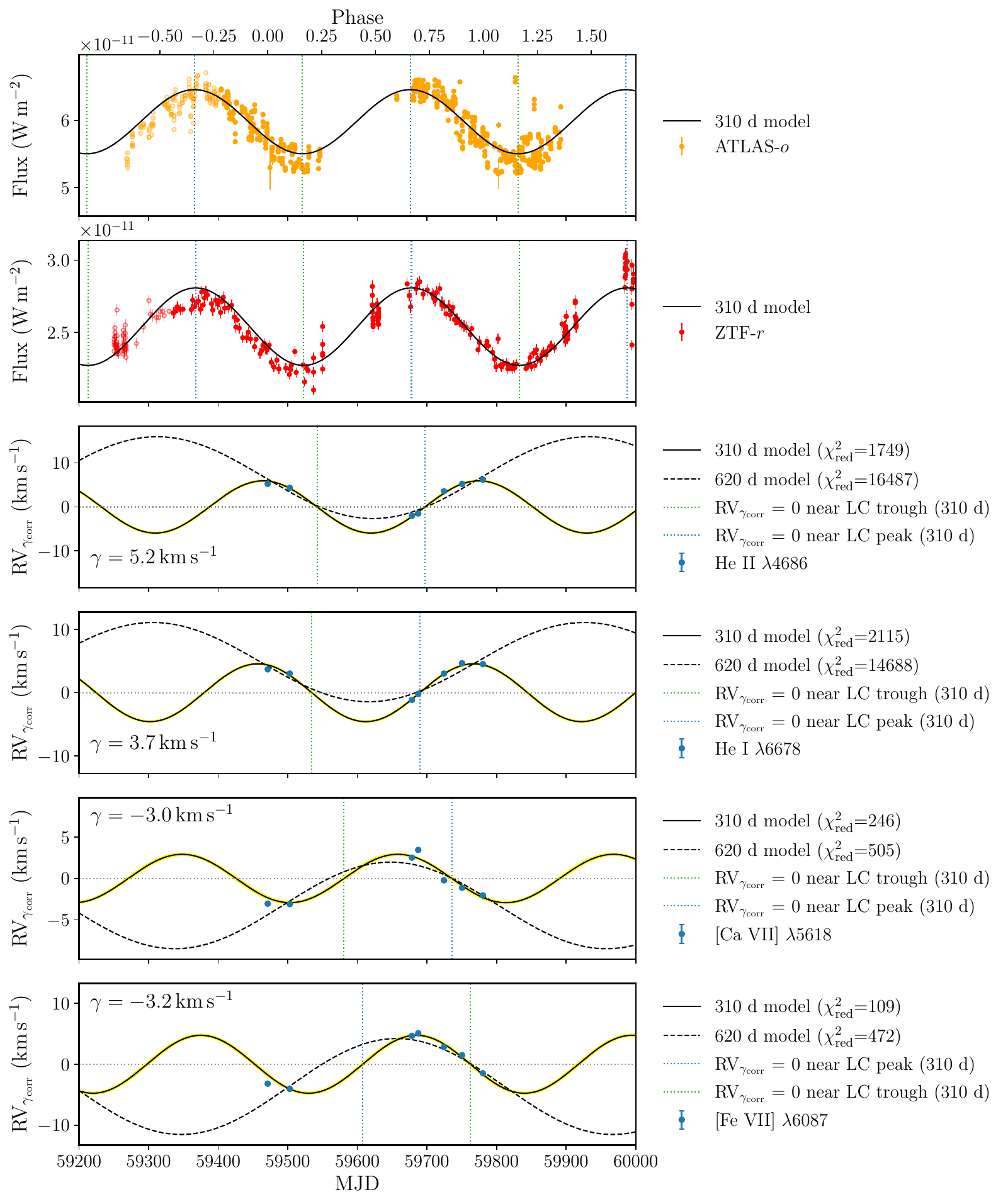}
\caption{Time-domain LCs and RVs of NaSt1. The top panels show ZTF-$r$ and ATLAS-$o$ photometry, and the lower panels present the $\gamma$-corrected RVs measured from He \ion{He}{2} $\lambda4686$, \ion{He}{1} $\lambda6678$, [\ion{Ca}{7}] $\lambda5618$, and [\ion{Fe}{7}] $\lambda6087$, where the systemic velocity ($\gamma$) has been subtracted from each dataset. Open symbols denote photometry previously reported by \citet{Lau_2021}, while filled symbols represent new observations presented in this work. Solid curves show sinusoidal models with fixed periods of 310 d, and dashed curves in the RV panels correspond to models with a period of 620 d. Vertical dotted lines indicate the offsets between RV$_{\rm corr}$ = 0 and the LC peaks/troughs, using the LC reference from the 310-day sinusoidal model. The thin yellow curves in the RV panels show 100 realizations drawn from the parameter uncertainties of the 310-day sinusoidal solution, illustrating the range of RV curves allowed by these uncertainties.}
\label{fig:lc_rv_binary}
\end{center}
\end{figure*}

We modeled the RV variations of each emission line with a sinusoidal function, assuming a circular orbit. The model parameters and their uncertainties were estimated using Markov Chain Monte Carlo (MCMC) sampling with the \texttt{emcee} package \citep{Foreman_2013}. Figure~\ref{fig:NaSt1_RV_Variation} shows the RV curves and their best-fit models for all 36 emission lines included in the period analysis. We did not fit the \ion{He}{2}$~\lambda4859$, H$\beta$~$\lambda4861$, and H$\alpha$~$\lambda6563$ lines because their profiles are contaminated by nearby spectral lines.
The period is $311 \pm 5$ d, where the quoted uncertainty represents the standard error of the mean. 
The standard deviation of the 33 period measurements in Table~\ref{tab:period_fit} is 27 d, corresponding to a $\sim$9\% variation, indicating scatter among the lines.
The mean period agrees with the 310 d photometric period reported by \citet{Lau_2021}. The derived periods, RV values, equivalent widths (EWs), and peak classifications for each emission line profile are listed in Table~\ref{tab:period_fit}.
Most lines vary in phase, apart from the high-ionization [\ion{Fe}{7}] and [\ion{Ca}{7}] lines, which are nearly $180^\circ$ out of phase from all other lines, as shown in Figures~\ref{fig:NaSt1_RV_Variation} and \ref{fig:lc_rv_binary}.
We will discuss the implications of these RV behaviors in Section~\ref{sec:discussions}. 



\begin{table*}[]
\centering
\caption{Periods derived from the CCF analysis, along with the RV amplitudes, EWs, and peak classifications for 36 emission lines of NaSt1. EW uncertainties are 1$\sigma$ standard deviations from seven epochs.}
\label{tab:period_fit}
\begin{tabular}{llcccc}
\hline
\hline
$\lambda_{\rm air}$ (\AA) & Spectral feature & Period (d) & RV amplitude (km\,s$^{-1}$) & EW (\AA) & Peak type \\
\hline
3889 & \ion{He}{1}      & 239.0 $\pm$ 1.0  & 1.81 $\pm$ 0.04 & 0.40 $\pm$ 0.07 & Single\\   
4658 & [\ion{Fe}{3}]    & 295.4 $\pm$ 2.2  & 5.57 $\pm$ 0.09 & 1.23 $\pm$ 0.16 & Double\\   
4686 & \ion{He}{2}      & 308.6 $\pm$ 0.3  & 5.95 $\pm$ 0.01 & 6.70 $\pm$ 0.91 & Single\\   
4701 & [\ion{Fe}{3}]    & 332.4 $\pm$ 3.5  & 6.29 $\pm$ 0.07 & 0.44 $\pm$ 0.07 & Double\\   
4713 & \ion{He}{1}      & 265.4 $\pm$ 0.2  & 5.19 $\pm$ 0.01 & 1.18 $\pm$ 0.13 & Single\\   
4859 & \ion{He}{2}      & 317.2 $\pm$ 3.0  & 4.05 $\pm$ 0.05 & 3.70 $\pm$ 0.53 & Single\\   
4861 & H$\beta$         & 308.6 $\pm$ 0.3  & 5.95 $\pm$ 0.01 & 3.56 $\pm$ 0.53 & Single\\   
4922 & \ion{He}{1}      & 307.3 $\pm$ 0.2  & 6.17 $\pm$ 0.01 & 2.15 $\pm$ 0.29 & Single\\   
5016 & \ion{He}{1}      & 292.6 $\pm$ 0.1  & 4.91 $\pm$ 0.01 & 7.10 $\pm$ 0.81 & Single\\   
5041 & \ion{Si}{2}      & 334.4 $\pm$ 3.7  & 3.14 $\pm$ 0.09 & 0.75 $\pm$ 0.13 & Double\\   
5048 & \ion{He}{1}      & 271.4 $\pm$ 2.1  & 6.59 $\pm$ 0.14 & 0.40 $\pm$ 0.08 & Single\\  
5056 & \ion{Si}{2}      & 309.2 $\pm$ 0.9  & 7.00 $\pm$ 0.07 & 0.93 $\pm$ 0.12 & Double\\   
5271 & [\ion{Fe}{3}]    & 346.2 $\pm$ 32.1  & 1.50 $\pm$ 0.13 & 1.62 $\pm$ 0.21 & Double\\   
5317 & \ion{Fe}{2}      & 313.9 $\pm$ 2.2  & 3.89 $\pm$ 0.04 & 0.99 $\pm$ 0.13 & Double \\  
5363 & [\ion{Fe}{2}]    & 321.0 $\pm$ 2.9  & 5.09 $\pm$ 0.05 & 0.56 $\pm$ 0.06 & Double\\   
5412 & \ion{He}{2}      & 292.5 $\pm$ 0.3  & 5.96 $\pm$ 0.02 & 4.53 $\pm$ 0.49 & Single\\   
5618 & [\ion{Ca}{7}]    & 337.0 $\pm$ 2.2  & $-$2.97 $\pm$ 0.03 & 2.89 $\pm$ 0.28 & Single\\   
5667 & \ion{N}{2}       & 365.0 $\pm$ 7.5  & 4.23 $\pm$ 0.11 & 0.66 $\pm$ 0.08 & Double\\   
5676 & \ion{N}{2}       & 331.2 $\pm$ 2.3  & 4.84 $\pm$ 0.09 & 2.38 $\pm$ 0.19 & Double\\   
5679 & \ion{N}{2}       & 320.0 $\pm$ 2.7  & 4.33 $\pm$ 0.10 & 0.57 $\pm$ 0.05 & Double\\   
5686 & \ion{N}{2}       & 319.1 $\pm$ 1.7  & 5.13 $\pm$ 0.11 & 1.83 $\pm$ 0.11 & Double\\   
5721 & [\ion{Fe}{7}]    & 314.7 $\pm$ 2.3  & $-$3.75 $\pm$ 0.07 & 1.37 $\pm$ 0.16 & Single\\   
5755 & [\ion{N}{2}]     & 300.9 $\pm$ 10.0  & 1.51 $\pm$ 0.16 & 23.78 $\pm$ 2.07 & Double\\  
5876 & \ion{He}{1}      & 290.8 $\pm$ 0.3  & 3.27 $\pm$ 0.01 & 111.77 $\pm$ 8.62 & Single\\  
6087 & [\ion{Fe}{7}]    & 335.4 $\pm$ 1.8  & $-$4.51 $\pm$ 0.04 & 4.19 $\pm$ 0.35 & Single\\    
6347 & \ion{Si}{2}      & 331.2 $\pm$ 5.4  & 3.81 $\pm$ 0.10 & 3.10 $\pm$ 0.18 & Double\\    
6372 & \ion{Si}{2}      & 314.0 $\pm$ 3.6  & 4.38 $\pm$ 0.13 & 5.07 $\pm$ 0.28 & Double\\    
6456 & \ion{Fe}{2}      & 319.8 $\pm$ 4.6  & 5.17 $\pm$ 0.09 & 2.28 $\pm$ 0.16 & Double\\    
6482 & \ion{N}{2}       & 367.8 $\pm$ 10.0  &  4.20 $\pm$ 0.10 & 1.51 $\pm$ 0.06 & Double\\    
6548 & [\ion{N}{2}]     & 305.2 $\pm$ 6.3  &  1.38 $\pm$ 0.15 & 10.07 $\pm$ 0.43 & Double\\   
6563 & H$\alpha$        & 328.4 $\pm$ 0.5  & 2.05 $\pm$ 0.01 & 136.90 $\pm$ 5.43 & Single\\  
6584 & [\ion{N}{2}]     & 294.5 $\pm$ 4.1  & 1.72 $\pm$ 0.15 & 31.27 $\pm$ 1.51 & Double\\   
6678 & \ion{He}{1}      & 308.1 $\pm$ 0.3  & 4.59 $\pm$ 0.01 & 170.61 $\pm$ 9.63 & Single\\  
7065 & \ion{He}{1}      & 320.8 $\pm$ 0.4  & 3.96 $\pm$ 0.01 & 531.71 $\pm$ 20.29 & Single\\ 
7282 & \ion{He}{1}      & 308.3 $\pm$ 0.5  & 5.06 $\pm$ 0.01 & 120.56 $\pm$ 9.52 & Single\\  
7468 & \ion{N}{1}       & 254.6 $\pm$ 4.4  & 3.08 $\pm$ 0.16 & 0.02 $\pm$ 0.01 & Double\\  
\hline
\end{tabular}
\tablecomments{Periods were not determined for \ion{He}{2}~$\lambda$4859, H$\beta$~$\lambda$4861, and H$\alpha$~$\lambda$6563 because their line profiles are contaminated by a nearby spectral line.}
\end{table*}

\subsection{Dust Parameter Estimation in the mid-IR}

Using the 1--5~$\mu$m spectrum, we corrected the observed fluxes for interstellar extinction, assuming $A_{\rm V} = 6.5$~mag and $R_{\rm V} = 3.1$ \citep{Crowther_1999,Fitzpatrick_1999}.
For wavelengths corresponding to $0.3 \leq 1/\lambda \leq 3.3~\mu{\rm m}^{-1}$, we adopt the extinction law of \cite{Fitzpatrick_1999}. At longer wavelengths ($\lambda > 3.3~\mu$m), we apply the near- and mid-IR extinction relations from \cite{Indebetouw_2005}, scaled to the adopted visual extinction. 
The extinction-corrected fluxes were computed as $F_{\rm dered} = F_{\rm obs}\,10^{0.4A_\lambda}$, and uncertainties were propagated accordingly. We flux-calibrated the extinction-corrected spectrum using the $K_{\rm s}$-band photometry from the Two Micron All Sky Survey (2MASS; \citealp{Skrutskie_2006}). We computed a synthetic $K_{\rm s}$-band flux by integrating the spectrum weighted by the $K_{\rm s}$ filter transmission curve. The synthetic flux density was then compared to the observed $K_{\rm s}$-band flux derived from the catalog magnitude of $K_{\rm s} = 6.87$ mag using the standard zero-point of 666.7 Jy. A multiplicative scale factor was applied to the spectrum to match the observed photometric flux. 

We modeled the mid-IR spectrum of NaSt1 using an optically thin dust emission model. The formulation of the dust flux follows \citet{Shahbandeh_2023, Tinyanont_2025}. The observed dust flux is given by
\begin{equation}
 F_{\rm dust}(\lambda) = \frac{B(\lambda, T_{\rm dust})\kappa(\lambda)M_{\rm dust}P_{\rm esc}(\tau)}{d^2}, 
\label{eq:observed_dust_flux}
\end{equation}
where $B$ is the Planck function, $d$ is the distance to the system, and $\kappa$ is the dust opacity \citep{Draine_1984, Laor_1993}. 

The mid-IR spectrum around 10~$\mu$m reported by \citet{Smith_2001} shows no silicate features, so we adopt carbonaceous dust in our modeling. 
We assume a representative grain size of $a = 0.1\,\mu$m. 
In the optically thin regime, the inferred dust temperature is only weakly dependent on the grain size for submicron grains \citep{Fox_2010, Sarangi_2022}. 
Assuming spherical symmetry, the escape probability $P_{\rm esc}(\tau)$ \citep{Cox_1969} is a function of the optical depth $\tau$,
\begin{equation}
P_{\rm esc}(\tau) = \frac{3}{4\tau}\left[ 1 - \frac{1}{2\tau^2} + \left( \frac{1}{\tau} + \frac{1}{2\tau^2} \right)e^{-2\tau} \right].
\label{eq:P_esc}
\end{equation}
For a uniform spherical dust distribution, the optical depth is
\begin{equation}
\tau(\lambda) = \rho R \kappa(\lambda) = \frac{3}{4}\frac{M_{\rm dust}}{\pi R^2}\kappa(\lambda).
\label{eq:optical_depth}
\end{equation}

We adopt a Gaia EDR3 distance of 3.0 $\pm$ 0.2 kpc based on a parallax of 0.33 $\pm$ 0.02 mas \citep{Gaia_2021}. The dust-emitting region has an angular Full Width at Half Maximum (FWHM) of $\sim$171 mas along the major axis in the $K_{\rm s}$ band \citep{Mauerhan_2015}, corresponding to a physical FWHM diameter of $\sim$500 AU ($R \sim 250$ AU) at this distance.

\begin{figure}[!hbt]
\begin{center}
\includegraphics[width=0.475\textwidth]{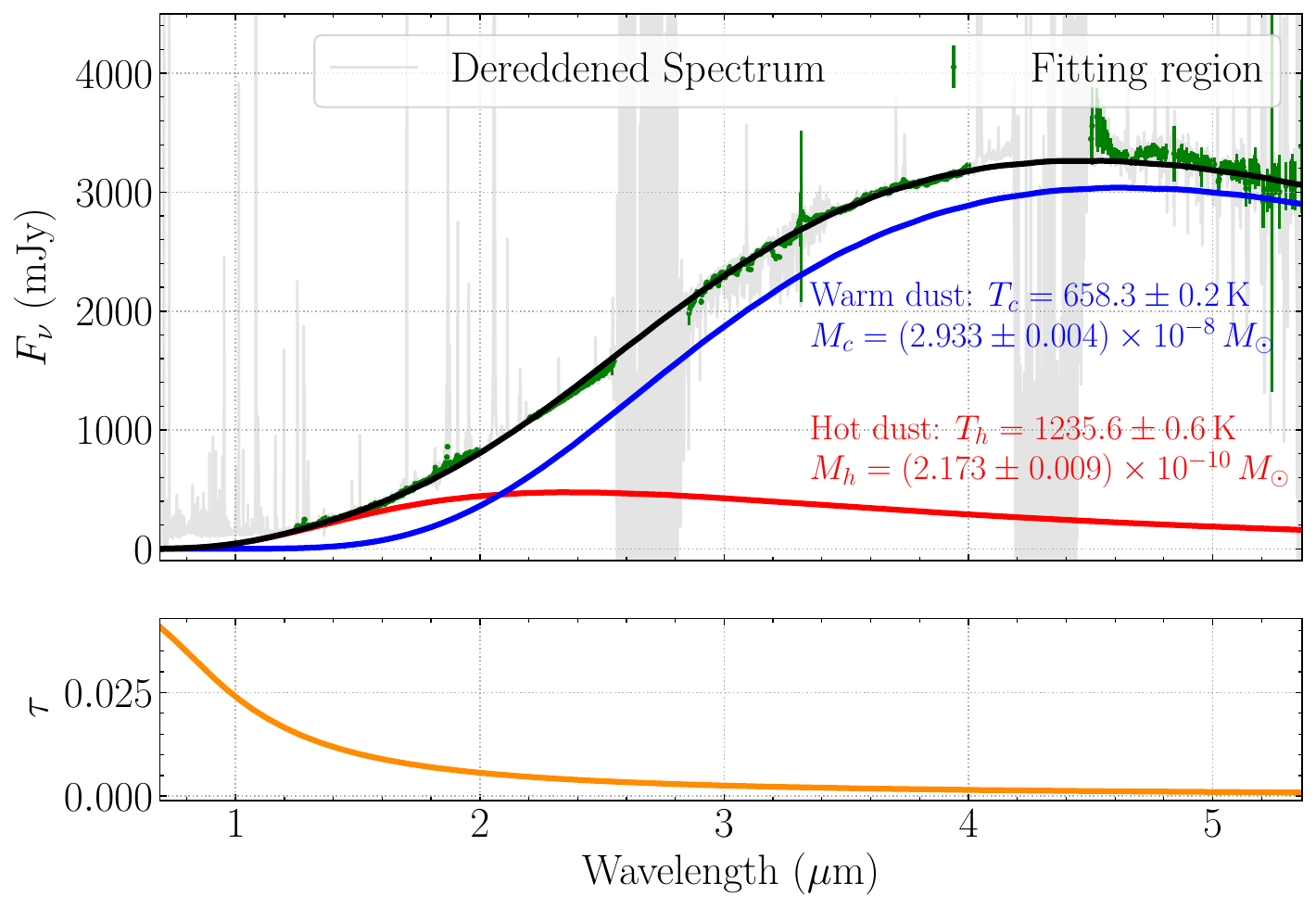}
\caption{Mid-IR spectrum of NaSt1 and the best-fitting two-component dust model, comprising hot and warm dust components constrained by MCMC fitting (top panel). The bottom panel shows the modeled optical depth, which remains below $\tau \sim 0.03$ across 1--5~$\mu$m and decreases toward longer wavelengths, consistent with optically thin emission.}
\label{fig:dust_temperature}
\end{center}
\end{figure}

We fit Equation~(\ref{eq:observed_dust_flux}) to the mid-IR spectrum using nonlinear least-squares minimization with \texttt{curve\_fit} from \texttt{scipy.optimize}, solving simultaneously for the dust temperature and dust mass. The model includes four free parameters: the temperatures ($T_{\rm h}, T_{\rm c}$) and masses ($M_{\rm h}, M_{\rm c}$) of the hot and cold dust components. Initial single-component dust models fail to reproduce the observed IR spectral energy distribution (SED). A two-component dust model provides a significantly improved fit.
The parameters and uncertainties of this two-component model were explored using MCMC, with the affine-invariant ensemble sampler \texttt{emcee} \citep{Foreman_2013} (as shown in Figure~\ref{fig:dust_temperature}). We find dust temperatures of $T_{\rm h} = 1235.6 \pm 0.6$  K and $T_{\rm c} = 658.3 \pm 0.2$ K for the hot and warm components. These values are consistent with the $\sim$2000 K and $\sim$700 K dust components inferred by \citet{Crowther_1999} from optical-IR spectrophotometry and IRAS photometry using multi-blackbody fits. The fitted dust masses are $M_{\rm h} = (2.173 \pm 0.009) \times 10^{-10}\,M_{\odot}$ for the hot component and $M_{\rm c} = (2.933 \pm 0.004) \times 10^{-8}\,M_{\odot}$ for the warm component. The quoted uncertainties are statistical only and do not include the dominant systematic uncertainty from the distance. The SED fit shows that the dust emission in NaSt1 is optically thin, with optical depths $\tau \ll$ 1. We note that in the optically thin regime, the assumed spherical geometry in the optical thickness no longer applies as $P_{\rm esc} \approx$ 1, and we see emissions from all dust grains.

\subsection{Spatially resolved kinematics of gas surrounding NaSt1}
\label{sec:kinematics_nast1}

\begin{figure*}[!ht]
\centering
\includegraphics[width=1\linewidth]{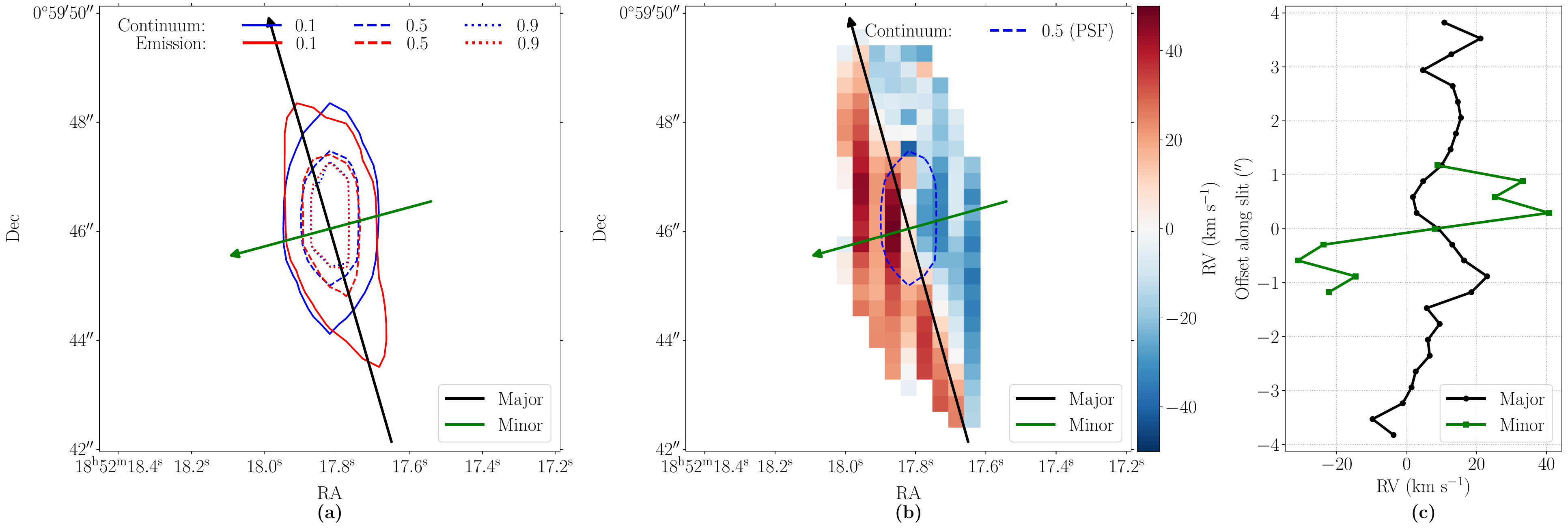}
\caption{
\textbf{(a)} Continuum (blue) and [\ion{N}{2}] $\lambda$6584 emission (red) along the projected major (black) and minor (green) axes. Contours are shown at 0.1, 0.5, and 0.9 of their respective peak intensities. 
\textbf{(b)} Radial velocity (RV) map of the [\ion{N}{2}] emission, with the same projected axes overplotted. The blue contour indicates the 0.5 continuum level, representing the point spread function (PSF) of an unresolved source. 
\textbf{(c)} RV profiles along the major (black) and minor (green) axes, showing peak-to-peak amplitudes of $\sim10$ km s$^{-1}$ and $\sim30$ km s$^{-1}$, respectively.
}
\label{fig:rv_map}
\end{figure*}

The IFU data provide spatially resolved spectroscopy, allowing us to investigate spatial extension in all emission lines in addition to the [\ion{N}{2}] $\lambda$6584 resolved in the HST narrow-band imaging \citep{Mauerhan_2015}. To determine which lines are spatially resolved, we compare the continuum and emission line flux distributions along the projected major and minor axes (Figure~\ref{fig:rv_map}a).
The continuum remains spatially unresolved. The \ion{He}{2} $\lambda4686$ and H$\alpha~\lambda6563$ profiles are marginally broader than the continuum but do not show clear spatial extension. In contrast, the [\ion{N}{2}] $\lambda6548$ and $\lambda6584$ lines exhibit significant extension beyond the continuum profile. The [\ion{N}{2}] $\lambda6584$ line shows the most significant spatial extension among all lines examined. This extension is consistent with the nebular structure previously resolved in HST imaging by \cite{Mauerhan_2015}.

We constructed RV maps for the resolved [\ion{N}{2}] lines. Rest wavelengths were converted from air to vacuum following \cite{Morton_1991}. At each spatial pixel, a Gaussian profile plus a constant continuum was fitted to the spectrum in the velocity space, and the Doppler velocity was computed from the centroid shift relative to the vacuum rest wavelength. 
The systemic velocity was defined as the median RV of all valid pixels within the emitting region and was subtracted from the maps.

Figure~\ref{fig:rv_map}b shows the RV  map of the [\ion{N}{2}] $\lambda6584$ line with blue- and redshifted emission separated along the projected major axis, indicating a resolved velocity gradient across the nebula. The [\ion{N}{2}] $\lambda6548$ line exhibits a similar velocity pattern, but with lower flux. The spatial extent and elongated morphology of the [\ion{N}{2}] emission are consistent with the disk-like structure resolved in the HST imaging \citep{Mauerhan_2015}. Furthermore, the RV gradient detected along our projected major axis, with an amplitude of $\sim$10 km s$^{-1}$ (Figure~\ref{fig:rv_map}c), is comparable to the gradient measured in their long-slit spectroscopy. We note that while the CSM is spatially extended, it is only resolved in the [\ion{N}{2}] $\lambda\lambda$6548, 6584 lines. The brightest part of the extended CSM from HST imaging is 2\arcsec along the major axis, so most flux falls into the 2\arcsec APF/Levy slit regardless of the PA. Further, the most extended regions exhibit relatively small RVs, so this does not significantly affect our RV measurements.


\section{Discussions} \label{sec:discussions}
\subsection{Binary nature of NaSt1}

The RVs derived from 36 emission lines yield a period of 
311 d, consistent with the LC period reported by \citet{Lau_2021}. 
To compare the proposed orbital-period scenarios, we evaluate the quality of the RV fits using the reduced chi-square statistic, 
$\chi_\nu^2 =
\frac{1}{\nu}
\sum_{i=1}^{N}
\frac{(O_i-C_i)^2}{\sigma_i^2}$,
where $O_i$ and $C_i$ denote the observed and model RVs, respectively, $\sigma_i$ is the formal RV measurement uncertainty, and $\nu=N-p$ is the number of degrees of freedom, with $N$ the number of observations and $p$ the number of free parameters.

\cite{Lau_2021} suggested that the LC variability might arise from ellipsoidal modulation in a binary system, in which case the orbital period would be twice the photometric period. 
Figure~\ref{fig:lc_rv_binary} compares sinusoidal models for RV with period equals to (310 d), and twice (620 d) that of the light curve. 
The 620 d model gives a factor of several larger reduced $\chi_\nu^2$ values than the 310 d solution. For the \ion{He}{2} $\lambda4686$ line, the best-fitting 620 d model yields $\chi_\nu^2 = 16487$, compared to $\chi_\nu^2 = 1749$ for the 310 d solution. The large reduced $\chi_\nu^2$ values indicate that the observed RV scatter is larger than expected from the formal RV uncertainties alone, likely due to intrinsic emission-line profile variability discussed in Section~\ref{sec:line_forming}. Comparable results are obtained for the other emission lines. 
Therefore, we disfavor ellipsoidal variation as the source of the variability in NaSt1.

Pulsations can also produce periodic photometric and RV variability, but they predict a different phase relation between the LC and RV. In pulsating stars, the RV is zero near the phases of minimum and maximum stellar radius, leading to a phase offset between the LC and RV variations. In contrast, the RV variations from most of the lines in NaSt1 occur in phase with the LC variations (Figure~\ref{fig:lc_rv_binary}). The RV variations exhibit time offsets relative to the LC, ranging from approximately $-$10 to $-$70 d for a 310 d period. The offset depends on the emission line, with \ion{He}{2} $\lambda4686$ and \ion{He}{1} $\lambda6678$ leading the LC by $\sim$53 and 71 d, respectively, while the higher-ionization lines [\ion{Ca}{7}] $\lambda5618$ and [\ion{Fe}{7}] $\lambda6087$ show smaller leads of $\sim$30 and 10 d, respectively.

This line-dependent behavior suggests that the emission does not arise from a single region. Higher-ionization lines originate closer to the central source and respond more promptly to the underlying variability, while lower-ionization lines arise at larger radii and exhibit correspondingly larger time offsets.
Such behavior is difficult to reconcile with stellar pulsations and instead favors an orbital origin (see Figure~3 of \citealt{Goldberg_2024}).
The RVs are measured from emission lines formed in the circumstellar environment rather than from stellar photosphere. The periodic modulation observed across multiple lines suggests that the variations are related to orbital motion in the binary system. This behavior supports the interpretation that NaSt1 is a binary system.

Although the formal RV uncertainties from the APF spectra are small, the measurements show scatter about the best-fitting sinusoidal curves, particularly for \ion{He}{1}, [\ion{Ca}{7}], and [\ion{Fe}{7}]. This scatter is likely astrophysical in origin. The potential origin of this scatter is discussed in Section~\ref{sec:profile_variability}.

\subsection{Line-forming regions}
\label{sec:line_forming}

\begin{figure*}[!hbt]
\begin{center}
\includegraphics[width=0.78\textwidth]
{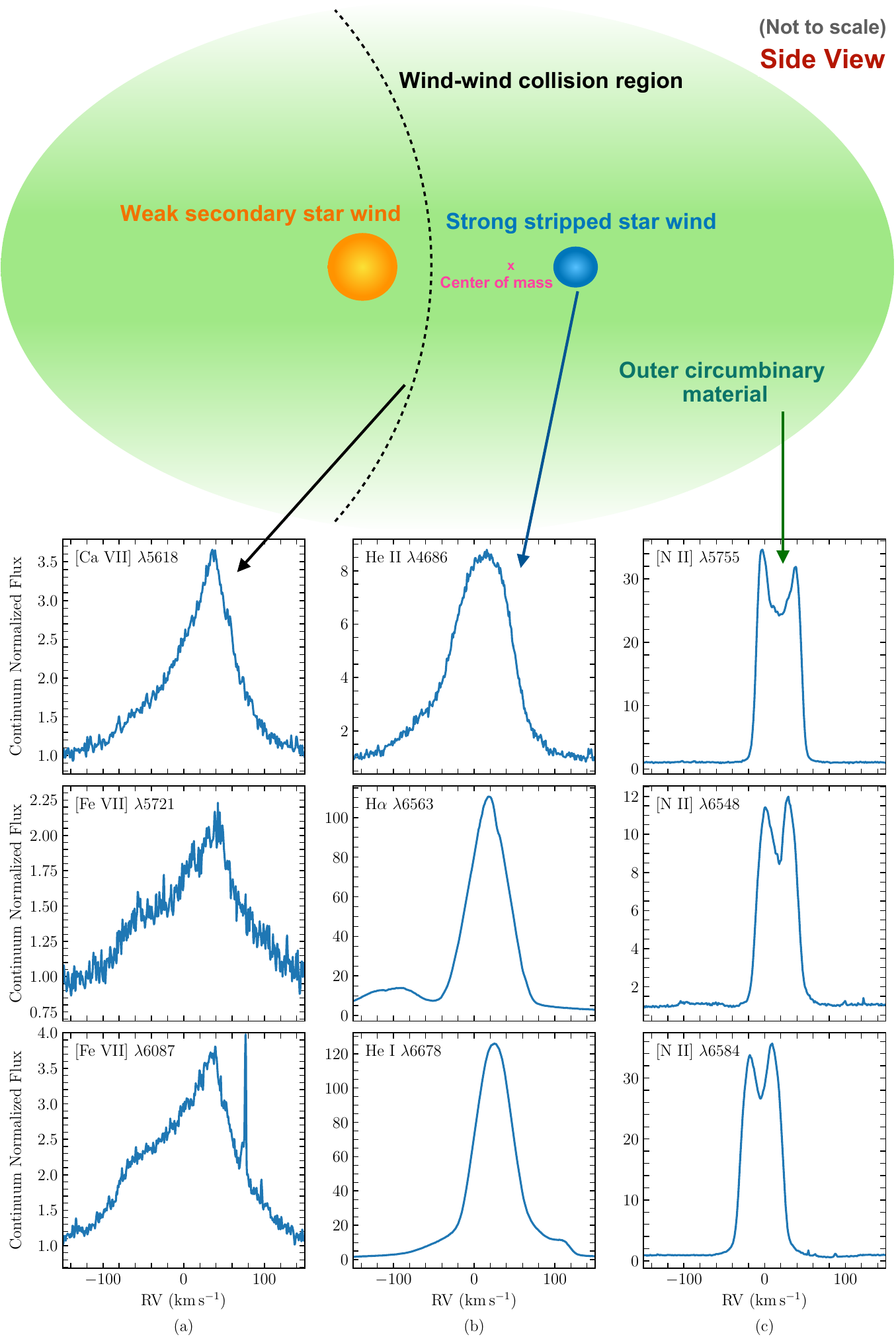}
\caption{Schematic illustration of the proposed circumstellar configuration of NaSt1. The stripped star drives a strong stellar wind, while the less massive companion produces a weaker wind, forming a wind–wind collision (WWC) region between the two stars. The blue region represents the inner circumbinary material (ICM), and the green region indicates the outer circumbinary material (OCM). Example emission-line profiles associated with these regions are shown in the lower panels: a) WWC ([\ion{Ca}{7}] $\lambda$5618, [\ion{Fe}{7}] $\lambda$5721, and [\ion{Fe}{7}] $\lambda$6087), (b) ICM (\ion{He}{2} $\lambda$4686, H$\alpha$ $\lambda$6563, and \ion{He}{1} $\lambda$6678), and (c) OCM ([\ion{N}{2}] $\lambda$5755, $\lambda$6548, and $\lambda$6584).}
\label{fig:NaSt1_Schematic_and_Spectra}
\end{center}
\end{figure*}

The emission-line spectrum of NaSt1 indicates contributions from multiple regions within the CSM. Based on their ionization states, line profiles, and RV phase behaviors, we divide the 36 emission lines into three groups. The first group corresponds to emission from the wind-wind collision (WWC) region. The second group traces the inner circumbinary material (ICM). The third group originates in the outer circumbinary material (OCM). Figure~\ref{fig:NaSt1_Schematic_and_Spectra} illustrates a schematic interpretation of the circumstellar environment of NaSt1 based on our spectroscopic results. Figure~\ref{fig:line_profiles_all} shows the line profiles of all 36 emission lines at all epochs (see Appendix~\ref{sec:spectroscopic_analysis}).

The WWC group consists of high-ionization lines, namely [\ion{Ca}{7}] $\lambda$5618 (ionization potential 127.2 eV), and [\ion{Fe}{7}] $\lambda\lambda$5721,~6087 ($\sim$125 eV). These lines exhibit single-peaked and asymmetric profiles. 
Their RV variations follow the same $\sim310$ d modulation with amplitudes of $\sim10$ km s$^{-1}$. All high-ionization lines exhibit RV variations that are nearly opposite in phase to those of all other emission lines (Figure~\ref{fig:NaSt1_RV_Variation}a). This behavior is explained if the high-ionization lines originate in the WWC region of a binary system. If the stripped primary drives a stronger wind than the secondary, the momentum balance places the WWC region closer to the secondary star (Figure \ref{fig:NaSt1_Schematic_and_Spectra}). 
Emission arising in this region traces the kinematics of the circumstellar gas rather than the orbital motion of the stars. The gas may have a different velocity component relative to the surrounding CSM, producing RV variations that appear in opposite phase compared to lines formed elsewhere in the system.
The high ionization energies indicate that these lines originate in a hot and energetic region, consistent with formation in the WWC zone observed in interacting massive binaries.

WWC regions are common in interacting massive binaries \citep{Stevens_1992}, including the WN5+O6 system V444 Cyg \citep{Rauw_2016}, the WC8+O7.5 III system $\gamma^2$ Velorum \citep{Zhekov_2007}, and the WN8+OB system WR~147 \citep{Henley_2005}. Their spectra are typically dominated by recombination lines (e.g., \ion{He}{2}) and high-ionization forbidden emission, up to, e.g., [\ion{Ca}{4}], [\ion{Ne}{3}], and [\ion{S}{4}]. In contrast, NaSt1 exhibits higher ionization ([\ion{Ca}{7}] and [\ion{Fe}{7}]), which have not been observed in other interacting massive binaries, indicating that the CSM around NaSt1 is unusually dense and close to the ionizing photon source.

The second group (ICM) consists of the hydrogen Balmer lines (H$\alpha$ $\lambda$6563 and H$\beta$ $\lambda$4861 (13.6 eV)) and helium emission lines, including \ion{He}{1} $\lambda\lambda$3889, 4713, 4922, 5016, 5048, 5876, 6678, 7065, and 7282 (24.6 eV), together with \ion{He}{2} $\lambda\lambda$4686, 4859, and 5412 (54.4 eV).
These lines exhibit single-peaked, more symmetric profiles and relatively small RV amplitudes of $\sim$5 km s$^{-1}$ (Figure~\ref{fig:NaSt1_RV_Variation}b). Their lower ionization energies ($\sim$10--55 eV) and weaker velocity variations suggest formation further out from the hotter WWC region. This is the region in the CSM that could be more influenced by the wind of the stripped primary star. 

The third group (OCM) includes metal lines such as 
[\ion{Fe}{3}] $\lambda\lambda$4658, 4701, 5271, 
\ion{Si}{2} $\lambda\lambda$5041, 5056, 6347, 6372, 
\ion{Fe}{2} $\lambda\lambda$5317, 6456, 
[\ion{Fe}{2}] $\lambda$5363, 
\ion{N}{2} $\lambda\lambda$5667, 5676, 5679, 5686, 6482, 
[\ion{N}{2}] $\lambda\lambda$5755, 6548, 6584,
and \ion{N}{1} $\lambda$7468. 
These lines exhibit double-peaked profiles and low ionization energies of $\sim$16--30 eV. Their RV variations follow the same $\sim$310 d modulation but with smallest amplitudes of $<$ 5 km s$^{-1}$ (Figure~\ref{fig:NaSt1_RV_Variation}c), suggesting that they are the furthest away from the central binary. 
The IFU data show that the [\ion{N}{2}] $\lambda\lambda$6548, 6584 emission lines are spatially resolved, further indicating that the OCM is the furthest out (see Section~\ref{sec:spatial_resolve}). Other lines in this group are not spatially resolved likely because of their weaker flux, but their similar line profiles and RV behavior suggest a common origin. 

\subsection{Line profile evolution in high-ionization emission lines}
\label{sec:profile_variability}

\begin{figure*}[!hbt]
\begin{center}
\includegraphics[width=1\textwidth]
{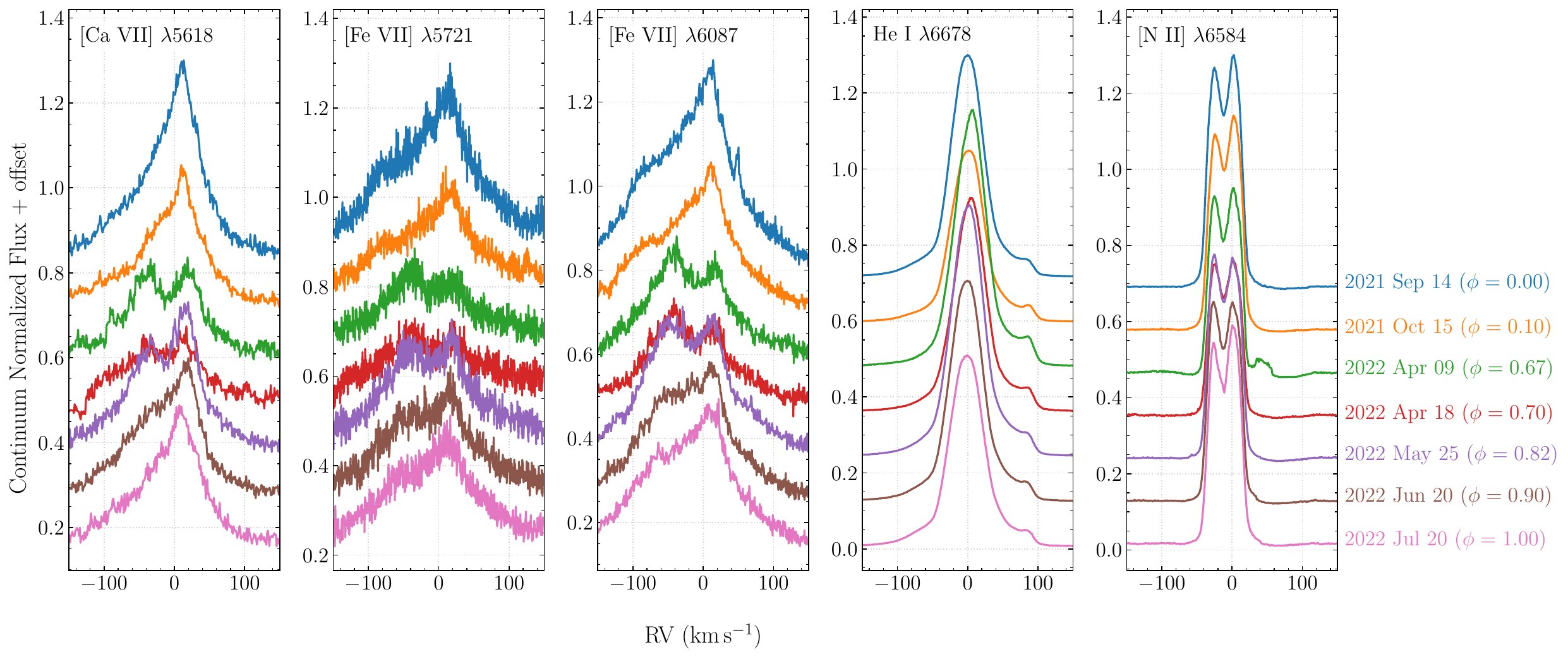}
\caption{Emission-line profiles from the APF high-resolution spectra of NaSt1 obtained over seven observing epochs, labeled with their corresponding observation dates and relative orbital phases ($\phi$). Each column shows the continuum-normalized flux as a function of radial velocity (RV) for a selected emission line, with the profiles vertically offset for clarity. The high-ionization forbidden lines ([\ion{Ca}{7}] $\lambda5618$ and [\ion{Fe}{7}] $\lambda\lambda5721,6087$) consistently show similar line-profile asymmetries and variability patterns across orbital phases, whereas the \ion{He}{1} $\lambda6678$ and [\ion{N}{2}] $\lambda6584$ profiles remain comparatively stable.}
\label{fig:line_profile_epochs}
\end{center}
\end{figure*}

Figure~\ref{fig:line_profile_epochs} compares select emission-line profiles across all seven observation epochs, with each profile labeled by its observation date and corresponding orbital phase ($\phi$). The high-ionization forbidden lines [\ion{Ca}{7}] $\lambda5618$ and [\ion{Fe}{7}] $\lambda\lambda5721,6087$ exhibit variations in both profile shape and relative line strength across the orbital cycle. The three high-ionization lines consistently exhibit similar asymmetric profiles and variability patterns at each epoch, suggesting a common origin within the same localized physical structure. In contrast, the lower-ionization lines, \ion{He}{1} $\lambda6678$ and [\ion{N}{2}] $\lambda6584$, remain comparatively stable throughout all epochs.

\begin{figure*}[!hbt]
\begin{center}
\includegraphics[width=1\textwidth]
{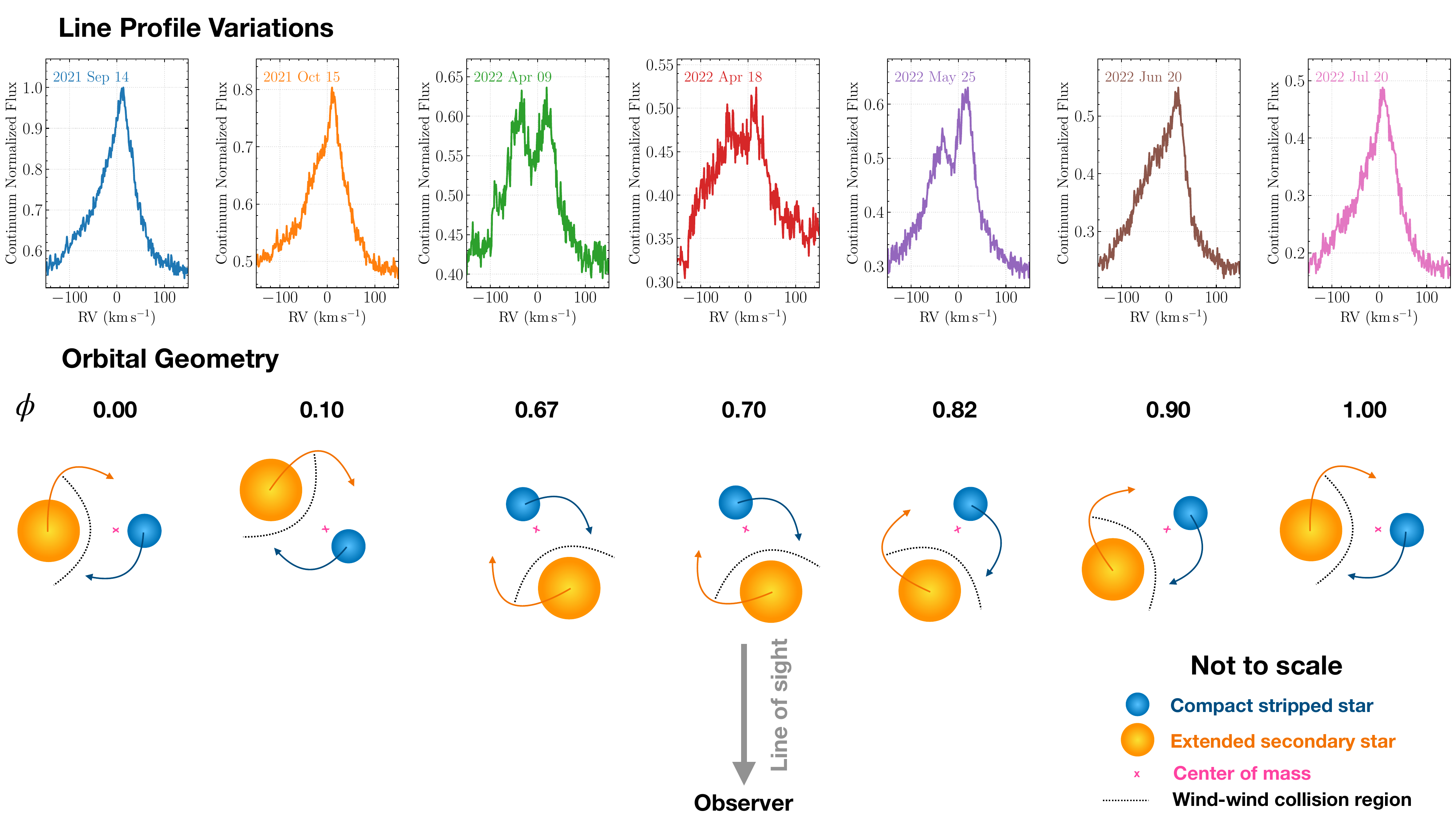}
\caption{Comparison of the observed [\ion{Ca}{7}] $\lambda5618$ emission-line profiles (top) with the proposed orbital geometry of NaSt1 (bottom). The orbital phases ($\phi$) are computed using the 310 day orbital period and correspond to the seven observing epochs. The observed variability is qualitatively consistent with orbital-phase-dependent changes in the projected geometry of the wind-wind collision (WWC) region. The schematic is not to scale.}
\label{fig:orbital_geometry}
\end{center}
\end{figure*}

The observed profile evolution is qualitatively consistent with the proposed orbital geometry shown in Figure~\ref{fig:orbital_geometry}. 
At phases of $\sim0.6$--$0.8$, the more extended secondary star obscures the WWC region, causing a flux deficiency near RV = 0 in the high-ionization lines. Similar geometry-driven flux deficits and line-of-sight absorption effects have been well-documented in other colliding-wind binaries, through both 3D hydrodynamical simulations \citep[e.g.,][]{Madura_2013, Richardson_2016} and observational studies \citep[e.g.,][]{Richardson_2017, Williams_2021}. 
The primary star is expected to obscure the WWC region around phase $\sim0.25$. However, the lack of observations at this phase prevents us from testing this effect.

Hydrodynamic instabilities and wind clumping within the WWC region may further enhance the observed variability \citep{Stevens_1992,Ignace_2009}. Because the RVs are measured using cross-correlation, these profile variations can shift the measured line centroids and contribute to the scatter around the best-fitting orbital solution. The orbital geometry shown in Figure~\ref{fig:orbital_geometry} is intended only as a qualitative illustration of this interpretation. A detailed hydrodynamic and radiative-transfer model of the WWC region is beyond the scope of the present work.

\subsection{Circumstellar medium expansion speed and mass-loss timescale}
\label{sec:spatial_resolve}

\cite{Mauerhan_2015} reported a peak-to-peak RV variation of $\sim$10 km s$^{-1}$ in the [\ion{N}{2}] line along the projected major axis of the nebula, based on long-slit spectroscopy. They noted that, if the CSM is an expanding disk, this value does not represent the maximum expansion velocity because the major axis tracks the transverse velocity. Higher velocity is along the projected minor axis, which traces radial motion.

Our IFU observations provide full two-dimensional kinematic information. We detect a clear RV pattern with blue- and redshifted emission clearly separated (Figure~\ref{fig:rv_map}b). Along the projected major axis, we measure a peak-to-peak velocity amplitude of $\sim10$ km s$^{-1}$, consistent with the previous long-slit measurement \citep{Mauerhan_2015}. Along the projected minor axis, the amplitude increases to $\sim30$ km s$^{-1}$, which traces the radial motion (Figure~\ref{fig:rv_map}c).

The HST imaging presented by \cite{Mauerhan_2015} shows that the nebula has a disk-like geometry tilted by $\approx 12 \pm 3^\circ$ from edge-on, corresponding to an inclination angle of $i = 78 \pm 3^\circ$. Using this inclination, we deproject the velocity along the minor axis and get a CSM expansion velocity of $v_{\rm exp} \approx 31$ km~s$^{-1}$. For a characteristic nebular radius of $R \sim 250$ AU (based on the Gaia distance and the angular size measured by \citealp{Mauerhan_2015}), the corresponding dynamical timescale is $t \sim R/v_{\rm exp} \approx 40$ yr. 
This timescale is consistent with the 1980s, when archival IR observations show long-term elevated flux in NaSt1 at a 0.5 mag level between 1--5 $\mu$m \citep{Lau_2021}, which could indicate a weak eruption.

\subsection{Comparison with RY Scuti}

RY Scuti is a massive interacting eclipsing binary surrounded by a compact nitrogen-rich nebula produced by binary mass transfer \citep{Smith_2002RY,Vince_2008}. The system contains an accretion disk and a circumbinary toroidal nebula, indicating ongoing mass exchange and mass loss.

NaSt1 shows several similar properties. The strong [\ion{N}{2}] emission, structured RV variations, and the expanding equatorial nebula revealed by the IFU data suggest a circumstellar environment shaped by binary interaction. The nebula around NaSt1 appears more obscured and less spatially resolved than that of RY Scuti.

Using the expansion velocity of $\sim$31 km s$^{-1}$ and a characteristic radius of $\sim$250 AU estimated above, the dynamical timescale of the NaSt1 nebula is $\sim$40 yr. 
For comparison, the ionized rings around RY Scuti have a dynamical age of $\sim$120 yr \citep{Smith_2002RY}, based on an expansion velocity of $\sim$45 km s$^{-1}$, which is moderately higher than that of NaSt1. 
This suggests that the circumstellar structure around NaSt1 may trace an even more recent episode of binary-driven mass loss.

\section{Conclusions} \label{sec:conclusions}

Our observations together support a binary interpretation of the NaSt1 system.
RVs measured from 36 emission lines show a periodic modulation with a period of $P = 310$ d, consistent with the previously reported photometric variability.
The phase relationship between the RV and LC variations is inconsistent with typical classical radial stellar pulsations and instead favors an orbital origin. While it might be possible to devise a non-standard pulsation scenario that reproduces the observed phase offsets, such models are not considered here.
The RV curves show two groups of emission lines varying in opposing phase. Most lines likely trace gas associated with the stripped primary wind, while the highest-ionization lines ([\ion{Fe}{7}] and [\ion{Ca}{7}]) originate in the WWC region located closer to the secondary star. The presence of a WWC region itself implies the interaction of two stellar winds and therefore provides strong evidence that NaSt1 is a binary system.

The emission lines originate from multiple regions within the CSM. High-ionization lines (e.g., \ion{He}{2} and [\ion{Fe}{7}]) arise in the WWC region and show the largest RV amplitudes ($\sim10$ km s$^{-1}$). Neutral lines (e.g., \ion{He}{1} and H lines) exhibit single-peaked profiles with smaller RV variations ($\sim5$ km s$^{-1}$) and trace the ICM. Metal lines such as [\ion{Fe}{3}], \ion{Si}{2}, \ion{Fe}{2}, and [\ion{N}{2}] show double-peaked profiles with radial velocity variations of $<5$ km s$^{-1}$ and likely originate in the OCM. These results indicate that the circumstellar environment of NaSt1 is stratified.

Mid-IR spectroscopy modeling reveals two optically thin dust components with temperatures of $\sim1230$ K and $\sim660$ K, consistent with a multi-component circumstellar environment. IFU spectroscopy spatially resolves the kinematics of the extended nebula in the [\ion{N}{2}] emission lines. The velocity gradient is $\sim10$ km s$^{-1}$ along the projected major axis and up to $\sim30$ km s$^{-1}$ along the minor axis, consistent with an inclined equatorial outflow. Adopting a disk inclination of  $i \approx 78^\circ$ \citep{Mauerhan_2015}, the deprojected expansion velocity is $v_{\rm exp} \approx 31$ km s$^{-1}$. For a characteristic radius of $R \sim 250$ AU, the corresponding dynamical age is $\sim40$ yr, indicating that the nebula was produced by relatively recent mass loss. Given this short dynamical age, future high-resolution imaging with HST should be capable of directly measuring the expansion of the [\ion{N}{2}] nebula, providing an independent constraint on the expansion velocity.
These results show that NaSt1 is a massive binary embedded in a structured CSM produced by a recent binary interaction, which may represent a Galactic analog for a SESN progenitor system in the midst of envelope stripping process. Continued observations will track its further evolution allowing us to directly observe this crucial and rare moment in massive stellar evolution.
\begin{acknowledgments}

This work is based in part on spectroscopic observations obtained with the Levy Spectrometer on the 2.4 m Automated Planet Finder (APF) telescope at Lick Observatory. We thank the staff of Lick Observatory for their support during the observations. This work also made use of archival photometric data from several public surveys, including the Zwicky Transient Facility (ZTF), the All-Sky Automated Survey for Supernovae (ASAS-SN), the Asteroid Terrestrial-impact Last Alert System (ATLAS), and Palomar Gattini-IR (PGIR).

This research has made use of the Keck Observatory Archive (KOA), which is operated by the W. M. Keck Observatory and the NASA Exoplanet Science Institute (NExScI), under contract with the National Aeronautics and Space Administration (NASA). Computational analyses were performed using the Chalawan High-Performance Computing (HPC) facility on the Castor cluster at the National Astronomical Research Institute of Thailand (NARIT). This research has made use of the NASA Astrophysics Data System (ADS).

This work is supported by the Fundamental Fund of Thailand Science Research and Innovation (TSRI) through the National Astronomical Research Institute of Thailand (Public Organization) (FFB690078/0269). This work contains data obtained with the Infrared Telescope Facility, which is operated by the University of Hawaii under contract 80HQTR24DA010 with the National Aeronautics and Space Administration. Some of the data presented herein were obtained at Keck Observatory, which is a private 501(c)3 non-profit organization operated as a scientific partnership among the California Institute of Technology, the University of California, and the National Aeronautics and Space Administration. The Observatory was made possible by the generous financial support of the W. M. Keck Foundation.
J.A.G. acknowledges financial support from NASA grant 23-ATP23-0070. The Flatiron Institute is supported by the Simons Foundation.
The authors wish to recognize and acknowledge the very significant cultural role and reverence that the summit of Maunakea has always had within the Native Hawaiian community. We are most fortunate to have the opportunity to conduct observations from this mountain.

\end{acknowledgments}

\vspace{5mm}
\facilities{Lick/APF, ZTF, ASAS-SN, ATLAS, PGIR}

\software{numpy~\citep{Harris_2020}, astropy \citep{Astropy_2013,Astropy_2018}, Matplotlib~\citep{Hunter_2007}, emcee~\citep{Foreman_2013}}


\appendix

\setcounter{figure}{0}
\renewcommand{\thefigure}{A\arabic{figure}}

\setcounter{table}{0}
\renewcommand{\thetable}{A\arabic{table}}

\section{Spectroscopic analysis}
\label{sec:spectroscopic_analysis}

We present the results of the high-resolution APF spectroscopic observations. 
Figure~\ref{fig:line_profiles_all} displays continuum-normalized emission-line profiles and their corresponding RVs. The profile includes hydrogen Balmer lines, \ion{He}{1}, \ion{He}{2}, and metal transitions such as \ion{Fe}{2}, \ion{Fe}{3}, \ion{N}{2}, [\ion{N}{2}], \ion{Si}{2}, [\ion{Ca}{7}], and [\ion{Fe}{7}]. The profiles exhibit a range of morphologies, including single- and double-peaked structures.

\subsection{Line Profile Variability over One Orbital Cycle}

\begin{figure}[!hbt]
\begin{center}
\includegraphics[width=1\textwidth]{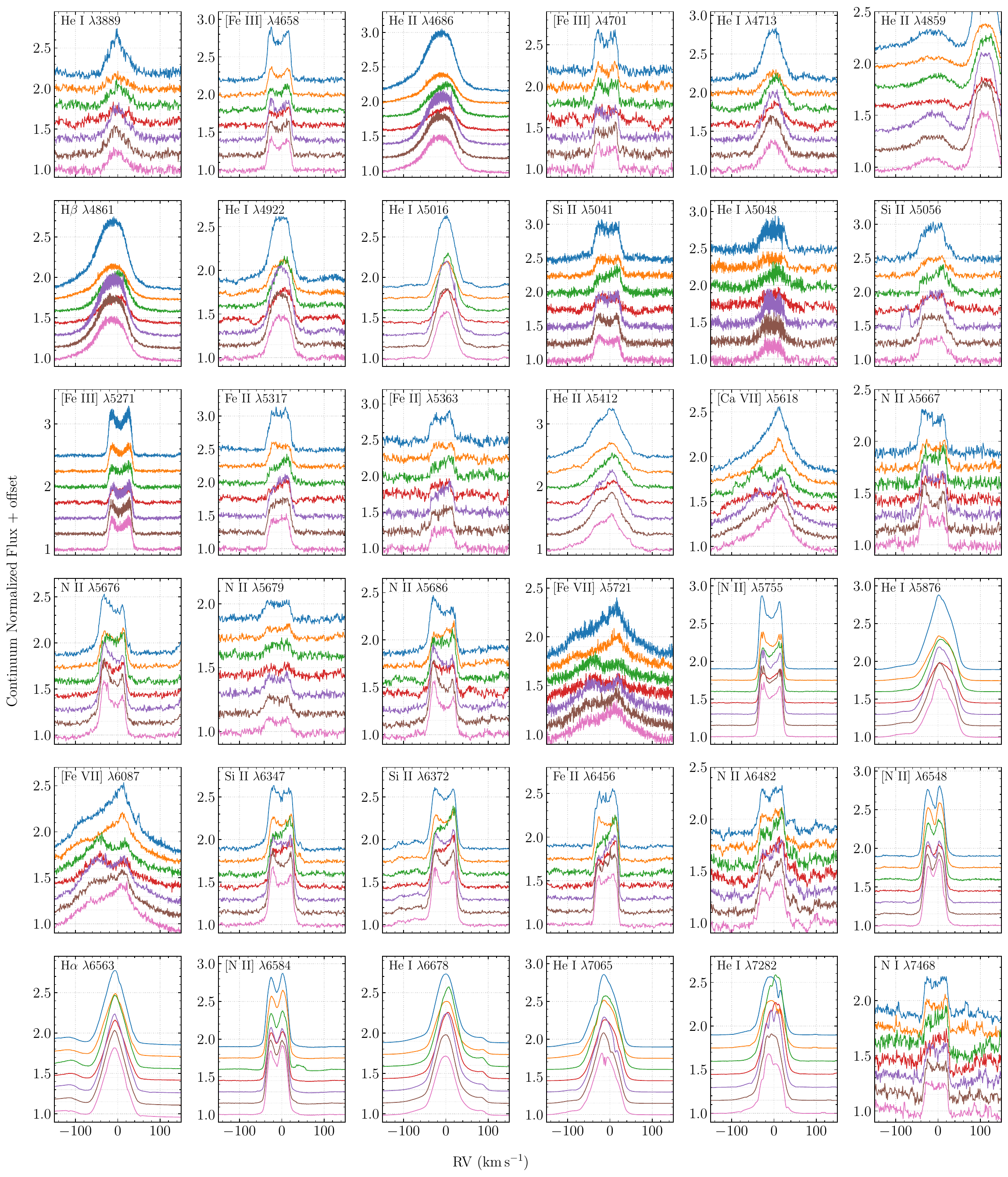}
\caption{Continuum-normalized line profiles of 36 emission lines observed over one complete orbital cycle of NaSt1. For each line, the seven spectra are displayed from top to bottom corresponding to observations obtained on 2021 September 14, 2021 October 15, 2022 April 9, 2022 April 18, 2022 May 25, 2022 June 20, and 2022 July 20. The line profiles are normalized to the peak intensity of each emission line and vertically offset for clarity.}
\label{fig:line_profiles_all}
\end{center}
\end{figure}

Figure~\ref{fig:line_profiles_all} presents the continuum-normalized profiles of 36 emission lines obtained over one complete orbital cycle. For each emission line, the seven spectra are displayed from top to bottom, corresponding to the observations obtained on 2021 September 14, 2021 October 15, 2022 April 9, 2022 April 18, 2022 May 25, 2022 June 20, and 2022 July 20. These observations span approximately one orbital period ($P \sim 310$ days).

The overall spectral morphology remains similar across the seven epochs, while individual emission lines exhibit different levels of variability. Representative low-ionization and recombination lines, such as \ion{He}{1}$~\lambda6678$ and [\ion{N}{2}]$~\lambda6584$, remain relatively stable, whereas high-ionization forbidden lines, including [\ion{Ca}{7}]$~\lambda5618$ and [\ion{Fe}{7}]$~\lambda5721$, $\lambda6087$, show significant changes in their line profiles. These epoch-dependent profile variations highlight the line-dependent variability in NaSt1 and motivate the discussion of the high-ionization line-forming regions in Section~\ref{sec:profile_variability}.

\subsection{Posterior distribution corner plots}
\label{sec:corner_plots}

\begin{figure}[!ht]
\begin{center}
\includegraphics[width=1.0\textwidth]{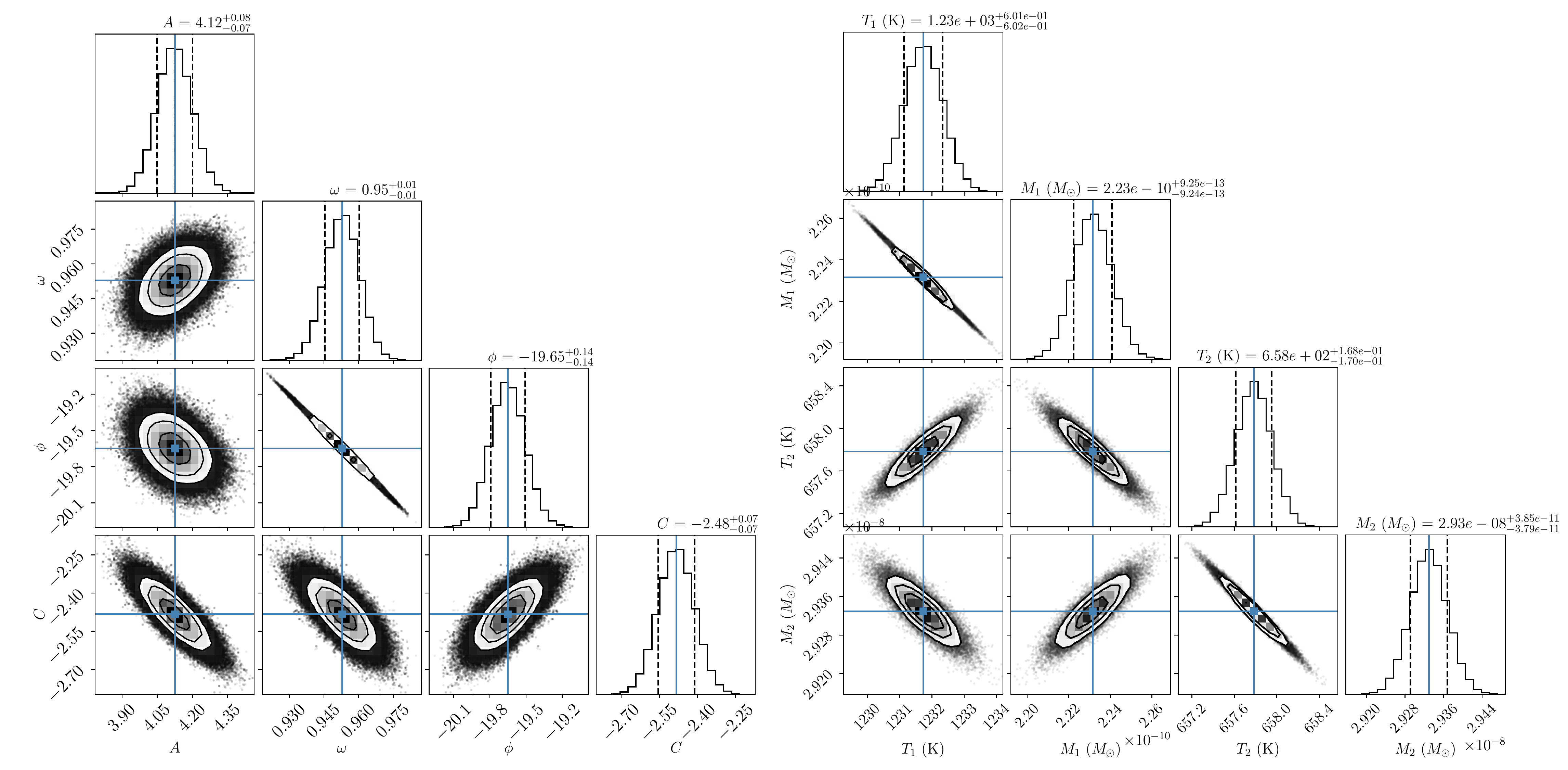}
\caption{Posterior probability distributions from the MCMC analysis. 
\textbf{Left:} Corner plot of the fitted parameters for the radial velocity (RV) model derived from the \ion{He}{1} $\lambda6678$ emission line, shown here as a representative example of the 36 emission lines analyzed.
\textbf{Right:} Corner plot of the parameters of the spectral energy distribution (SED) model, including the temperatures and masses of the two dust components. The dashed lines indicate the 16th, 50th, and 84th percentiles of the posterior distributions.
}
\label{fig:corner_HeI6678}
\end{center}
\end{figure}

We used the MCMC sampler implemented in \texttt{emcee} \citep{Foreman_2013} to explore the posterior distributions of the fitted parameters. 
Figure~\ref{fig:corner_HeI6678} presents the corner plots for the orbital period fitting (left) and the SED dust model (right). 

For the period fitting, we initialized 50 walkers and ran the chains for 10,000 steps, discarding the first 3,000 steps as burn-in. Convergence was assessed through inspection of the chains and autocorrelation analysis. 

For the SED fitting, we initialized 60 walkers and ran 10,000 steps, discarding the first 1,000 steps as burn-in. Convergence was similarly verified by examining the chain behavior and autocorrelation times.

\bibliography{NaSt1}{}
\bibliographystyle{aasjournalv7}

\end{document}